# TOWARDS A SOFTWARE PRODUCT SUSTAINABILITY MODEL

Coral Calero[1], Mª Ángeles Moraga[1], Manuel F. Bertoa[2]

[1]*Instituto de Tecnologías y Sistemas de la Información*

*University of Castilla-La Mancha. Spain*

Coral.Calero@uclm.es; MariaAngeles.Moraga@uclm.es

[2]*Departamento de Lenguajes y Ciencias de la Computación*

*University of Malaga*

*Malaga, Spain*

bertoa@lcc.uma.es

1. Introduction

   The necessity to adapt current products and services into a way of working environmentally friendly is already a social and economic demand. Although the GreenIT can be considered a mature discipline, software sustainability, both in its process and its use, has not begun to be a topic of interest until the last few years. In this sense we think is fundamental to define what we consider that is software sustainability and how to evaluate it properly.

   Sustainable software development refers to a mode of software development in which resource use aims to meet product software needs while ensuring the sustainability of natural systems and the environment and Sustainability of a software product can be defined as the capacity of developing a software product in a sustainable manner (Calero et al., 2013).

   As remarked in Calero et al. (2013) software sustainability is a way to improve a software product, being part of its quality and being related to non-functional requirements (requirements that constrain or set some quality attributes upon functionalities, Glinz, 2007).

2. Modeling software product sustainability

   As we have remarked, our point of start is that sustainability is part of a software product quality. ISO/IEC 25000 is a series of standards specific for System and software product Quality Requirements and Evaluation, namely SQuaRE (ISO/IEC 25000, 2010). In the ISO/IEC 25010 division, three quality models are defined (ISO/IEC 25010 (2010). In fig. 1 the targets of the quality models and the related entities are shown.

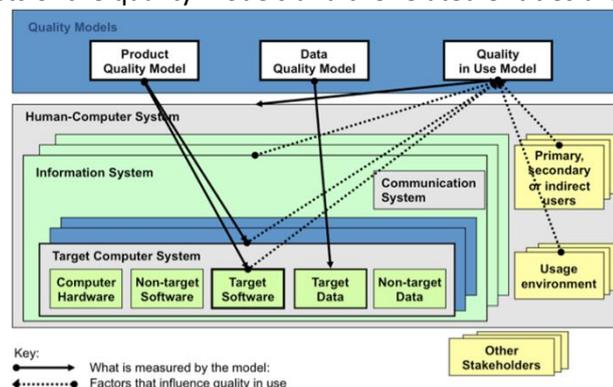

Fig. 1. Targets of quality model

From this figure we can observe that product quality is reflected on the Product quality model which is related to the target software. In the standard, the product quality model is composed of eight characteristics each one subdivided into several subcharacteristics (see fig. 2).

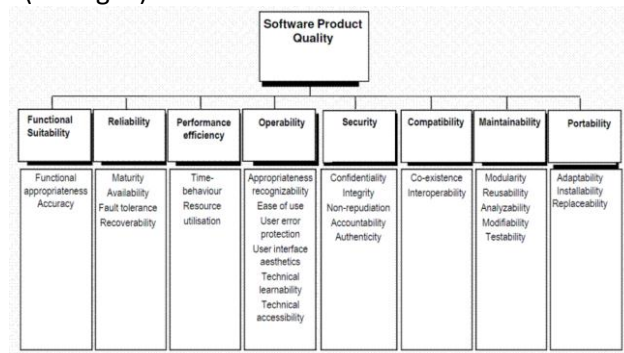

Fig. 2. Software product quality model in ISO/IEC 25010

As is stated in the standard, "the product quality model is useful for specifying requirements, establishing measures, and performing quality evaluations. The quality characteristics defined can be used as a checklist in order to ensure a comprehensive treatment of quality requirements, thus providing a basis that can be used to estimate the consequent effort and activities that will be needed during systems development. The characteristics in the product quality model are intended to be used as a set when specifying or evaluating software product quality".

So, if we want to consider sustainability as part of the quality model, it is necessary to define it together with its subcharacteristics.

For doing it, we must take into account that when a software product is being developed, its sustainability can be considered from two points of view.

First we must ensure that the software product is energy-efficient when it works, using the resources in the most appropriate manner. In Calero et al (2014) the authors propose the following subcharacteristics for sustainability, which would fit with this "short-term" point of view:

- **Energy consumption**. Degree to which the amounts of energy used by a software product when performing its functions meet requirements.
- **Resource optimization**. Degree to which the amounts and types of resources used by a product when performing its functions meet sustainability requirements. As in the standard, authors consider that resources can include: other software products, the software and hardware configuration of the system, and materials (e.g. print paper, storage media).

On the other hand we must ensure that the software product will endure over time, being only required its replacement if adapting it to the new circumstances is very difficult to achieve. We refer to this as **Perdurability**.

The idea of making a software perdurable is to achieve a software product lasting in time, that is modifiable, reusable, ie those aspects that make the software developed lasts time and is able to adapt to change without losing its functionality or other features related to its quality.

In order to define perdurability we need first to identify what it must consider and, for doing it, we are going to use the standard (ISO:25010, 2010). In the standard, there are 3 characteristics that could be related with what we are looking for:

**Reliability**. Degree to which a system, product or component performs specified functions under specified conditions for a specified period of time

- **Maturity**. Degree to which a system meets needs for reliability under normal operation

- **Availability**. Degree to which a system, product or component is operational and accessible when required for use
- **Fault tolerance**. Degree to which a system, product or component operates as intended despite the presence of hardware or software faults
- **Recoverability**. Degree to which, in the event of an interruption or a failure, a product or system can recover the data directly affected and re-establish the desired state of the system

**Maintainability**. Degree of effectiveness and efficiency with which a product or system can be modified by the intended maintainers

- **Modularity**. Degree to which a system or computer program is composed of discrete components such that a change to one component has minimal impact on other components
- **Reusability**. Degree to which an asset can be used in more than one system, or in building other assets
- **Analysability**. Degree of effectiveness and efficiency with which it is possible to assess the impact on a product or system of an intended change to one or more of its parts, or to diagnose a product for deficiencies or causes of failures, or to identify parts to be modified
- **Modifiability.** Degree to which a product or system can be effectively and efficiently modified without introducing defects or degrading existing product quality.
- **Testability**. Degree of effectiveness and efficiency with which test criteria can be established for a system, product or component and tests can be performed to determine whether those criteria have been met

**Portability.** Degree of effectiveness and efficiency with which a system, product or component can be transferred from one hardware, software or other operational or usage environment to another

- **Adaptability**. Degree to which a product or system can effectively and efficiently be adapted for different or evolving hardware, software or other operational or usage environments
- **Installability**. Degree of effectiveness and efficiency with which a product or system can be successfully installed and/or uninstalled in a specified environment
- **Replaceability**. Degree to which a product can be replaced by another specified software product for the same purpose in the same environment

However, if we look in detail these sub-characteristics and their definitions, we can discard many of them for not being related to long-term issues (the ones we are looking for). As a result, we obtain a final set of characteristics that could be related to perdurability. These characteristics must be taken into account when defining the Perdurability characteristic: **Reusability, Modifiability** and **Adaptability**.

These characteristics will not be used exactly as they are defined for defining the perdurability but they will be used as a basis for doing it and taking into account the objective of the perdurability.

Therefore, we can define the **perdurability** as the degree to which a software product can be modified, adapted and reused in order to perform specified functions under specified conditions for a long period of time.

Once we have identified and defined the subcharacteristics of the sustainability, from the both points of view described, we could add to the software product quality model proposed by the ISO/IEC 25010 (2010) a new characteristic for sustainability defined as shown in Figure 3:

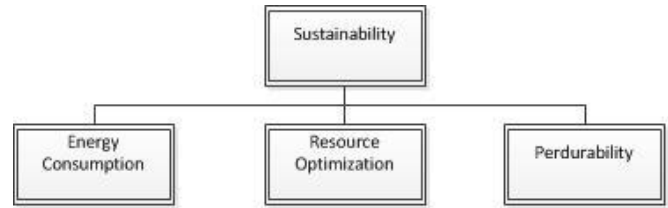

Fig. 3. The sustainability characteristic

3. Conclusions

In this paper we have presented the importance of the software sustainability. We have presented, from the assumption that sustainability is part of the quality of a software product, a specific model for software sustainability that can be added to the software product quality model from the standard ISO/IEC 25010 (2010).

From the definition of software sustainability and its characteristics it will be possible to incorporate the sustainability in the development of a software product, in the form of non-functional requirements and ensure that the final products are environmentally friendly.

Also from the definition of the sustainability characteristic it will be possible to define measures and indicators for the sustainability of a software product, being able to use them to evaluate, to detect weaknesses or to improve the sustainability of a software product. All these are our future lines of work and research.

4. Acknowledgment

This work is part of the GEODAS-BC (TIN2012-37493-C03-01), the PEGASO (TIN2009-13718-C02-01) and Sistemas Inalámbricos de Gestión de Información Crítica (TIN2011-23795) projects funded by the Spanish Ministerio de Economía y Competitividad and by FEDER (Fondo Europeo de Desarrollo Regional).

5. Referencias